# TASC-1D-cSi: a simulation tool to scrutinize the thermal impacts on the performances of crystalline silicon solar cells

Olivier Dupré, Mohamed Amara, and Rodolphe Vaillon

Université de Lyon, CNRS, INSA de Lyon, CETHIL – UMR 5008, F-69621 Villeurbanne, France

Abstract: The capabilities of a simulation tool for the in-depth analysis of the thermal impacts on the performances of solar cells are described. TASC-1D (Thermal Analysis of Solar Cells - 1D, version cSi) solves the coupled electrical, radiative and thermal transport problems for a crystalline silicon cell, as a function of irradiation and thermal conditions. In addition to the electrical outputs that are obtained with the existing simulation tools, it provides the cell equilibrium temperature as well as the spatial and spectral distributions of many relevant quantities. The physical modeling and associated numerical solution techniques are summarized. Result cases with a prescribed cell temperature demonstrate that the code performs well, is able to provide valuable information on the quantum efficiency and power loss mechanisms, and is capable of handling diffuse irradiations. The intricate correlation between thermal sources and optical-electrical losses is discussed. The other cases include the solution of the heat transfer problem and analyses are conducted on the cell operating temperature and efficiency as a function of the thermal conditions. This simulation tool is likely to allow new optimizations of photovoltaic cells that include a thermal criterion in addition to the optical and electrical criteria.

## I. INTRODUCTION

Among the scientific topics related to the optimization of the performances of solar cells, those regarding the optical, electrical and thermal losses are still a key issue. The first are addressed through the search of means for harvesting as much as possible of the sun's radiation over its whole spectral extent and the second by avoiding the recombination of the photogenerated electrical charges. As for the thermal effects, it is well known that an increase of the cell temperature leads to a drop in photo-conversion efficiency [1]. The review on the thermal performance of Si and GaAs solar cells exposed by Radziemska [2] discusses the main parameters affected by temperature. It is obvious that the dependence on temperature of the conversion efficiency is a function of the photovoltaic (PV) system and the operating conditions, such as solar irradiation and wind conditions. To deal with these parameters, it might be convenient to use correlations, such as those listed and discussed in the review of Skoplaki and Palyvos [3]. Correlatively, this means that the prediction of the operating temperature of a solar cell or an ensemble of cells is mandatory.

Analytical expressions of this temperature for various PV installations have been derived, as summarized in [4]. While they are well suited to be used in larger scale system analyses, they in fact hide a lot of complex phenomena. Even the implicit correlations, involving parameters themselves depending on temperature, cannot allow performing a thorough examination of the thermal impacts on solar cell performances. A more elaborate approach is thus to build a thermal model for solar cells. As an example, Jones and Underwood [5] proposed a time-dependent thermal model for a PV module, taking into account convection and radiation exchanges in addition to the electrical power output. While giving a good estimation of the time varying operating temperature of the cell, the intricate connections among thermal, electrical and spectral radiative phenomena cannot be analyzed yet with this kind of approach.

For a sharp understanding of the transport physics, dedicated simulation codes are required. The recent review by Altermatt [6] includes a historical overview on the developments of models for numerical device simulation of crystalline silicon solar cells. The list is not repeated here, however one can quote DESSIS and PC1D (a set of references can be found in [6]) as among the well-known computer programs simulating the electrical behavior of various solar cells or PV devices. But if these codes take into account the dependence of some parameters on temperature, they sometimes do not solve the heat transfer problem (the cell temperature is prescribed) and, in any case, none of them can catch the detailed coupling mechanisms involved among spectral radiation, thermal and electrical transport phenomena going on within the cell, in relation with the operating conditions (parameters of the cell, thermal and irradiation conditions). In order to tackle this issue, we started with the development of a computer program solving these coupled problems for a one-dimension representation of a crystalline silicon homo-junction. The basic equations, methods of solution, results and associated analyses were introduced with extensive explanations in a previous paper [7]. While we actually used the well-established models for the transport of electrical charge carriers in semiconductor devices, our contribution regarded the modeling of the spectral radiative transfer for both the collimated and diffuse components of irradiations and of heat transfer within the cell. The objective of the present paper is to introduce the simulation tool we have developed to be able to perform in-depth thermal analyses of solar cells. Specific uses of this computer code will be made to illustrate how it may help improving our understanding of known effects or possibly to come across with new ones.

The paper is organized as follows. The simulation code (named TASC - 1D for Thermal Analysis of Solar Cells) is described in section 2 in its version for crystalline silicon cells (TASC-1D-cSi). The problem under consideration, main elements of the theoretical formulations and numerical solution techniques are summarized. The succeeding section is dedicated to the delivery of results for a series of selected situations with the aim at addressing specific issues. After a description of additional validation tests of the code, we give an insight into its capabilities and present some results about the thermal behavior of c-Si cells.

## II. THE SIMULATION CODE: MAIN ELEMENTS OF THE MODELING AND NUMERICAL METHODS

### A. Problem description and main assumptions

A single N on P homo-junction without encapsulation but possibly with an anti-reflection coating (ARC) is considered (Figure 1, where the relative scales are not realistic). The cell dimensions in $y$ and $z$ directions are assumed very large compared to the thickness of the cell so that properties are only functions of the $x$ coordinate. The steady-state 1-D thermal - radiative - electrical simulation of the solar cell includes the parameters and applies to the operating conditions given hereafter. The cell is subdivided into a N-doped (emitter) region and a P-doped (base) region with thicknesses $w_N$ and $w_P$, respectively. At the front and back surfaces, the following boundary conditions (BCs) hold. Electrical BCs affect the charges created within the cell that flow into a circuit with a load of voltage ($V$). The electrical carriers may recombine at the front and back surfaces with given velocities (respectively $S_h$ and $S_e$ for holes and electrons). Radiative BCs encompass the solar and surrounding irradiations. At the front surface, solar radiation hits the cell with given proportions of collimated (with a polar angle $\theta_c$) and diffuse isotropically distributed rays with properly defined spectral fluxes per unit surface area. A surrounding diffuse isotropic irradiation can be taken into account at the front and/or the back surface via a blackbody intensity spectral distribution calculated at the temperature ($T_{ext,f}$ / $T_{ext,b}$) of the front/ back external medium. Surfaces are assumed perfectly smooth. Optical reflection at the front external medium - semiconductor interface can be lowered by the addition of thin layers of dielectric materials to build a so-called anti-reflection coating. Thermal BCs chiefly consist in defining heat exchange coefficients ($h_f$ and $h_b$), which together with the previously introduced external medium temperatures, stand for any kind of thermal exchanges (the coefficient is very large to model a prescribed temperature or zero for an insulated wall; otherwise it comes from conductive or natural - mixed - forced convective values or formulas). Despite their expected smaller effect, surface recombinations of charges are also incorporated in the thermal BCs.

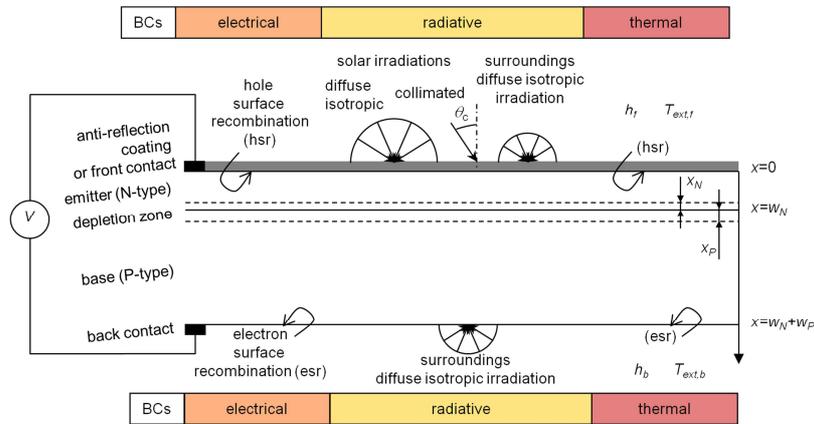

Fig. 1. Schematic view of the problem.

Properties of the cell depend on the semiconductor (crystalline silicon in this paper) and the doping levels (respectively donor and acceptor densities $N_d$ and $N_a$). The set of required electrical, radiative and thermal properties can be found in the seminal paper [7] and consequently details are not repeated here. However some changes were made and are reported briefly in the appendix. Whenever needed and available in the literature, any dependence on temperature of the aforementioned properties is properly considered.

By comparing dimensions of the cell with the mean free paths of each carrier (phonons, electrons and holes, photons), it can be assumed that heat and electrical charge transport are diffusive processes while the Radiative Transfer Equation for unpolarized radiation adequately describes the propagation of electromagnetic wave energy within the cell. The semiconductor material is absorbing and emitting. Reflections and transmissions at interfaces are governed by Snell's and Fresnel's laws.

With the aforementioned conditions, the simulation code solves the coupled radiation, heat and electrical carrier transport problems within the solar cell and provides as outputs the cell temperature distribution together with the photovoltaic conversion performances as well as a lot of optical, electrical and thermal information. A summary of the solution methods employed is given hereafter.

*B. Solution methods*

To evaluate in details the thermal impacts on the solar cell performance characteristics, the coupled electrical, thermal and radiation transport problems are to be solved. An overview of the algorithm is provided in Figure 2. Details on the equations and solution techniques can be found in [7] and the references quoted therein. First of all, the main inputs are loaded (marker M01): they consist of the boundary conditions and properties that do not depend on temperature. Consecutively, in accordance with all the spectral distributions of radiative properties, the spectrum is divided into chosen frequency bands (M02) spanning from 2.00 $10^{13}$ to 1.13 $10^{15}$ Hz, which corresponds in vacuum to the wavelength interval 0.2 – 15.0 µm. An initial temperature distribution (uniform) within the cell ($T_{init}$) is set to prescribe a value for the thermal conductivity (M03) whereas the internal heat sources are set to zero (M03). Thus, with the defined thermal BCs, the heat transfer problem is solved (M04) using the Thomas' algorithm solution of the matrix writing of the spatially discretized form of Fourier's law. From this, we get a first temperature profile within the cell (M05) and all thermo-dependent properties can be updated (M06).

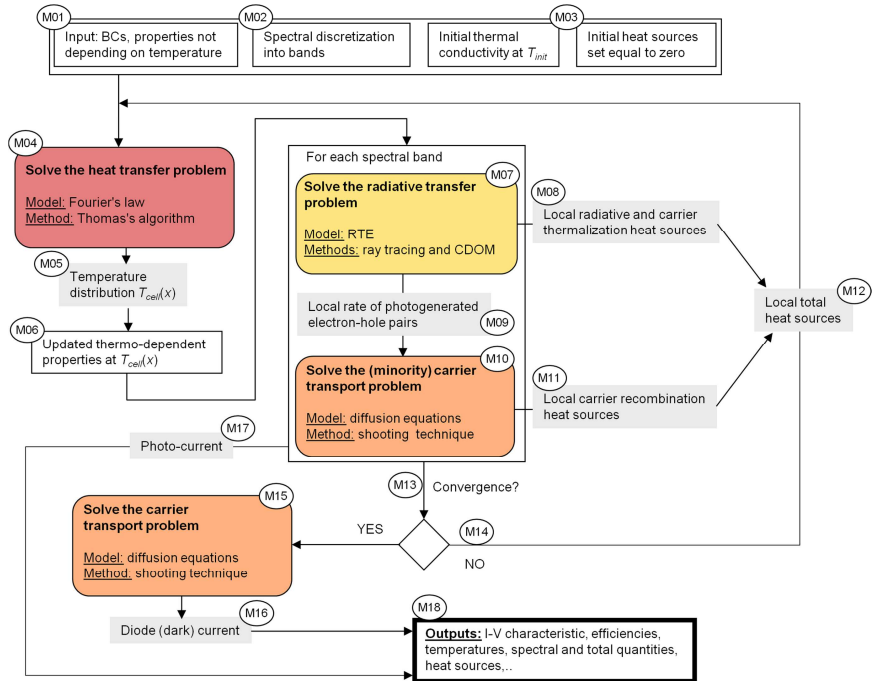

Fig. 2. Solution algorithm of TASC-1D.

For each band of the spectral discretization, the radiative transfer problem is solved (M07). A ray tracing technique is used for the collimated component of the solar irradiation and takes into account the multi-reflections within the cell. As for the diffuse component, a Composite Discrete Ordinates Method provides the local band integrated intensities at every location in the cell. Use of this technique requires spatial and directional discretizations. For the latter one, two subsets of discrete directions must be defined in accordance with the refraction and total reflection cones at the cell - external medium interface. The main calculated quantity is the -spatially- local band integrated incident radiation power. It is used to evaluate two of the local heat sources: the first one derived from the quasi-instantaneous thermalization of the photo-generated electrical charges and the other one being the balance between radiation locally emitted (at the local temperature) and absorbed then converted into heat (M08). It also gives the local rates of photo-generation of electron-hole pairs (M09). They are inserted into the minority carrier diffusion equations that are solved separately for holes (N-Region) and electrons (P-region), using the so-called shooting technique (M10). It outputs the minority carrier profiles used to calculate the local heat sources caused by the recombinations of carriers (M11). Once the radiative transfer and minority carrier transport problems are solved for all spectral bands, the summation of band integrated sources provides the local total (integrated on the whole spectral range of interest) heat sources (M12).

As we deal with coupled phenomena, an iterative process is required to converge towards stabilized values of thermal, radiative and electrical quantities. Thus, a convergence test is applied to decide whether a new iteration has to be run or not (M13). It consists in checking the thermal budget of the cell: the calculated total internal heat source for the whole cell has to be balanced by the fluxes at the boundaries. When the convergence criterion is not satisfied, a new iteration has to be implemented (M14), thus getting new solutions of the heat transfer, radiative transfer and minority carrier transport problems. When convergence is eventually reached, then the final temperature profile is at our disposal for terminal calculations. The dark (diode) current is estimated as a function of the load voltage (M15), using the same equations and techniques as when the cell is under illumination, except that boundary conditions for minority carrier diffusion equations are modified to model the thermally created majority carriers swept by the electric field through the depletion region towards the region where they are in a minority. This gives the diode current (M16) to be subtracted to the photocurrent. The latter was calculated during the last iteration (M17) from the addition of minority carrier concentration gradients in each doped region at the edge of the depletion region with the charges photo-generated and collected within the depletion region.

At last, a large set of outputs (M18) is available for subsequent analyses. Among these data we can quote: the current-voltage characteristic from which the maximum power and photo-conversion efficiency can be inferred, the internal and external quantum efficiencies, the spectral response, other spectrally varying quantities such as the heat sources, as well as their distribution within the cell, the contributions of the different parts of the cell to the photocurrent, and of course, the temperature distribution within the cell.

Table 1
CELL CHARACTERISTICS AND BOUNDARY CONDITIONS

| Cell thickness (μm) | $w_N + w_P = 250$ ; $w_N = 0.3$ | |
|---|---|---|
| Doping density (cm⁻³) | $N(x) = \begin{cases} 1.10^{18} & x \leq w_N \\ 1.10^{16} & w_N \leq x \leq w_P \end{cases}$ | |
| ARC | SiN ; $w_{ARC} = 90$ nm | |
| **Boundary conditions** | **front** | **back** |
| Electrical | $S_h = 100$ m s⁻¹ | $S_e = 100$ m s⁻¹ |
| Thermal | $h_f = 5 \ 10^4$ W m⁻² K⁻¹ | $h_b = 5 \ 10^4$ W m⁻² K⁻¹ |
| | $T_{ext,f} = 25$°C | $T_{ext,b} = 25$°C |
| Radiative | AM1.5D solar spectrum + diffuse blackbody radiation at 25 °C | Diffuse blackbody radiation at 25 °C |

The previous explanations about the algorithm and numerical techniques solving the problem illustrate the rather complexity of the resulting simulation code. The choice of the spectral and spatial discretizations is of paramount importance to correctly represent the spectral and spatial variations of all the different parameters (solar spectrum, refractive indices of silicon and the ARC materials,...). The new version of the code uses an exponential spatial meshing in order to accurately describe the exponential absorption of the solar flux. Since the thermal problem doesn't require much spatial accuracy because the temperature in the cell is in most cases uniform, a different meshing is used to solve it. Each modeled phenomenon has its own solution techniques (shooting method, ray tracing,...) that were carefully validated in [7]. Several budgets are used to monitor the power conservations. Heat, solar irradiation and global radiation power balances are checked for the converged quantities. The solar irradiation power balance consists in checking on each spectral band that the incoming sun's radiation is fully distributed in different terms: the reflected part at the front interface, the fraction that crosses the cell, the part that is transformed into heat through thermalization, surface and volume recombinations, and the fraction that is electrically collected. As for the cell global radiation power balance, it is worth noticing that it involves 13 different terms. A presentation of the other features that have been modified or added to the original version of the code described in [7] is proposed in the appendix (mobilities, intrinsic carrier concentration, ARC reflectivity calculations, lifetimes). In the following, we present some results from TASC-1D-cSi to give an insight into its capabilities.

## III. RESULTS AND ANALYSES

The objective of the present section is to put forward different results of the TASC-1D-cSi simulation tool and to present several analyses. Unless otherwise specified, the calculations are made for a basic configuration described below. The cell parameters, electrical / thermal / radiative boundary conditions and main transport properties for holes and electrons are summarized in Table 1 (the full list of properties is given in [7] and in the appendix). Results reported hereafter were computed using 119 spectral bands, 10000 meshes including 9000 where the exponential discretization is applied, and 120 discrete angles respectively for the spectral, spatial and angular discretizations. With these calculation parameters, thermal, interband absorption and radiative power budgets were found to be respectively less than 0.1%, 0.1% and 0.3%. A special attention should be paid to the thermal and radiative boundary conditions detailed here. The used high unrealistic heat transfer coefficients enable to set artificially the cell temperature equal to the external temperature. In section 3.3 we will use these coefficients to model a wide range of real PV device outdoor condition scenarios.

### A. Validations

As spectral quantities were not assessed originally in [7], an additional test is proposed hereafter. Spectral variations of the Internal and External Quantum Efficiencies (IQE and EQE) are calculated and compared with PC1D's results. Since PC1D does not solve the heat transfer problem and cannot handle diffuse irradiation, TASC-1D calculations are made with a prescribed cell temperature of 25 °C and using the reference spectrum AM1.5D that corresponds to a 1000.4 W.m⁻² direct irradiation. The resulting curves displayed in Figure 3 are fully satisfying since they are superimposed. It proves that the methods used in TASC-1D are sound.

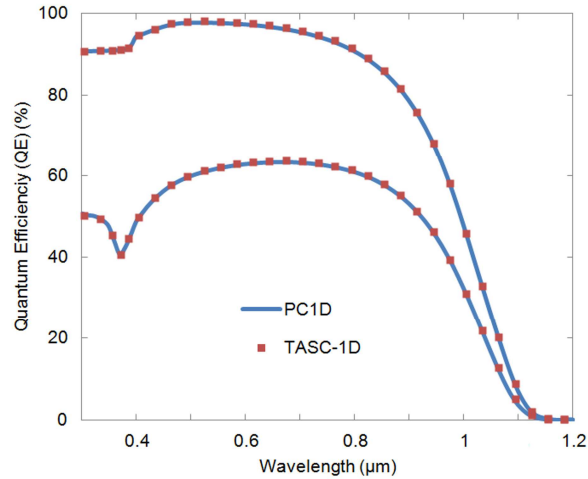

Fig. 3. IQE and EQE obtained with PC1D (blue lines) and TASC-1D (red squares).

Since TASC-1D-cSi is aimed at analyzing the thermal behavior of solar cells, the variation of electrical outputs as a function of temperature is assessed against existing data. The temperature of the cell is prescribed in the simulation code and varies from 25 to 75 °C with a 10 °C step. The resulting *J-V* and *P-V* characteristics are given in Figure 4. The slight increase of the short-circuit current density $J_{sc}$, the substantial drop of the open-circuit voltage $V_{oc}$ and the resulting decrease of the maximum power density output $P_{max}$ are fully consistent with known trends. The calculated temperature coefficients $\dfrac{1}{J_{sc}}\dfrac{dJ_{sc}}{dT}$, $\dfrac{1}{V_{oc}}\dfrac{dV_{oc}}{dT}$ and $\dfrac{1}{P_{max}}\dfrac{dP_{max}}{dT}$ respectively 0.0004, -0.0040 and -0.0047 compare quite well with typical values found in the literature (for example [2,3]).

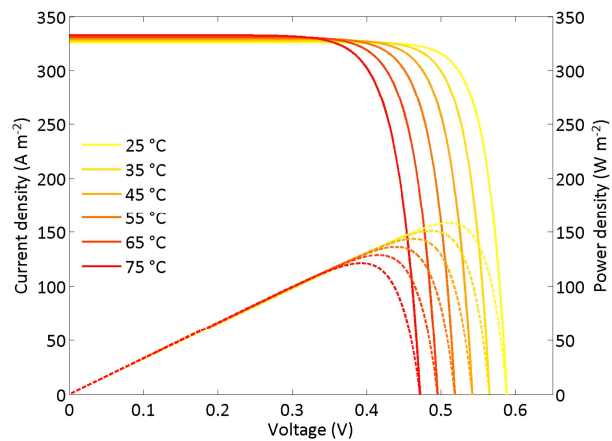

Fig. 4. J-V and P-V characteristics at different temperatures.

### B. An insight into some capabilities of TASC-1D-cSi

#### 1) Quantum efficiency loss mechanisms

Like any simulation tool of PV cells, TASC-1D-cSi can give the External and Internal Quantum Efficiencies (EQE and IQE) of the cell. But one of its originalities is that it provides a synthetic visualization to observe in a glance which mechanisms limit the photon conversion, on which wavelengths intervals and in which proportions (Figure 5).

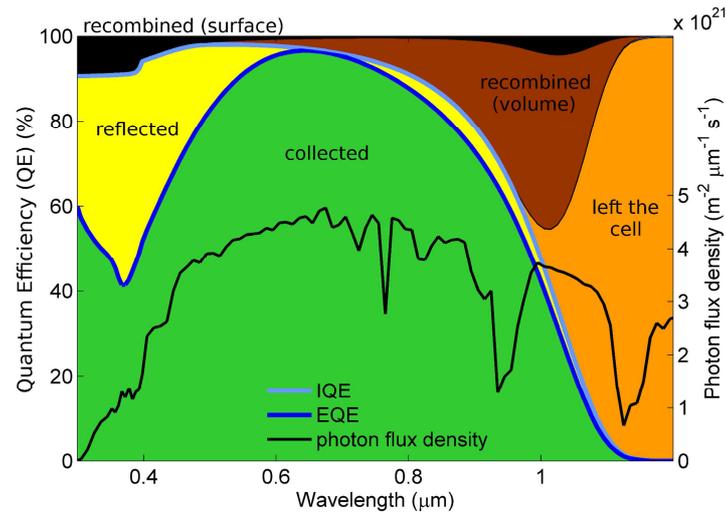

Fig. 5. Quantum efficiency loss mechanisms.

The first optical loss is due to the front interface reflection and is pictured by the yellow area between the EQE and IQE lines (in dark blue and light blue, respectively). One can notice the variation around 0.3 µm that corresponds to a peak in the refractive index of silicon. Another remarkable feature is that the reflection is almost zero around 0.65 µm. This is due to the thickness of the Anti-Reflection Coating (ARC) chosen in order to minimize the reflection at the wavelength where there is the largest number of incident photons. The second optical loss corresponds to the photons which leave the cell unabsorbed. Given the cell thickness and the operating temperature considered in the computations, the corresponding region - colored in orange -, starts from a little bit before 1 µm and becomes the main cause of QE losses, as a result of a decrease of the interband absorption coefficient with increasing wavelength. This flux leaving the cell might seem large but it is the case only because the enhanced reflection that back contacts would cause was not included in this modeling. The last contribution to optical losses is the photon absorption by the free carriers and the lattice. It is depicted by the tiny blue area almost invisible in the upper right corner between the brown and the orange areas. Its little size confirms the standard assumption that this mechanism is negligible in this spectral region.

All remaining photons create electron-hole pairs in the cell that may not all be collected: this gives rise to the so-called electrical losses. The amount of photogenerated electrons or holes that recombine at the surfaces is represented by the black area. This area clearly shows two maximums that are due to the recombinations at the front and the back surfaces. The addition of a back surface field in the modeling would likely decrease the second peak. The volume recombination depicted by the brown area logically increases with wavelength since more charges are generated far from the depletion zone and are therefore more likely to recombine before reaching it. It decreases after about 1 µm because the previously mentioned optical losses increasingly prevent the photons from creating charges.

The visualization of these different contributions is a direct mean to observe how cell parameters such as cell thickness, electrical carrier lifetime, or surface recombination velocity, affect the PV cell performances. To figure out the consequences of QE losses on the incoming radiation power that can be converted into electrical power, the black line is added to provide the spectral photon flux density of the considered solar irradiation. At this stage, the thermal losses are hidden thus an analysis of power loss mechanisms is required to highlight the thermal impacts.

*2) Power loss mechanisms*

Another originality of TASC-1D-cSi lies in its ability to determine precisely how the incident power is transformed on each spectral band (Figure 6). The fractions of incident power dissipated by the optical and electrical mechanisms described in the previous paragraph are depicted on this graph using the same color code. One can notice that the power lost through surface recombination is very low especially at short wavelengths while Figure 5 shows that this phenomenon has an important impact on quantum efficiency. This is explained by the thermalization process represented by the red area. The energy of a photon is inversely proportional to the wavelength but photogenerated charges can only retain a fixed amount of energy (determined by the band gap) while the remaining energy is thermalized (i.e. dissipated into heat in the cell).

When carriers recombine in volume or at surfaces, they convert their energies into heat. By looking at the sum of the red, the brown and the black areas, one can fully measure the extent of the fraction of the incident power that is dissipated into heat in a standard silicon solar cell. This observation emphasizes the importance of a thermal design for PV systems. Not only this energy is lost to the creation of electricity, it has another negative impact on the efficiency of the system. It corresponds to a heat source within the cell that, together with the thermal boundary conditions, will drive the cell temperature above the ambient temperature thus leading to a drop in efficiency.

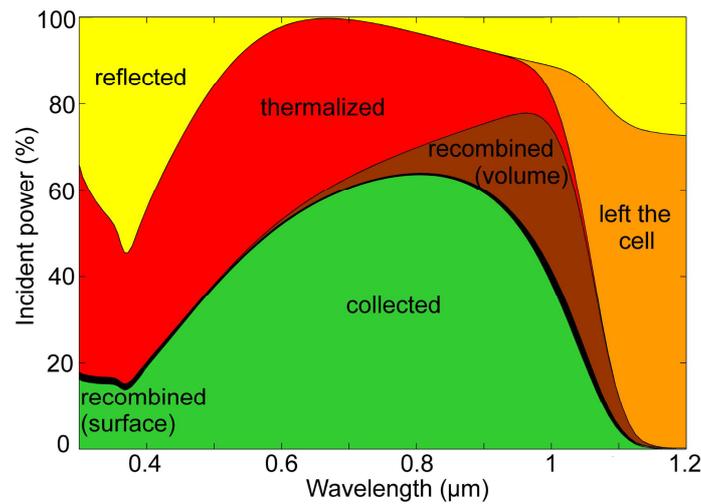

Fig. 6. Power loss mechanisms.

One can see on Figure 6 the percentage of the incident flux that is electrically collected and that can be used to generate an electric power at every wavelength (green colored area). It can be concluded that the ideal optimum spectrum (for a PV system with these characteristics operating at 25°C) would be a monochromatic irradiation at 805 nm.

In real conditions, we have to deal with the solar spectrum so it is important to consider the spectral variations and the directional nature of the incident flux. Figure 7 shows the power distribution in the case of the standard parameters given in Table 1 but considering both the direct and the diffuse components of the AM1.5 spectrum. This figure is meaningful about the capabilities of TASC-1D: it depicts how the irradiation fluxes – solar and from the surroundings – are spectrally distributed. The fluxes reflected and leaving the cell correspond to the optical and radiative losses. The collimated and diffuse component contributions are discernible. A special attention will be paid to the handling of the diffuse irradiations in the next section, with an emphasis on the importance of taking into account the infrared part of the radiation spectrum. The three internal heat sources due to thermalization, surface and volume recombination of photogenerated electrical carriers can be spectrally quantified. The remaining part corresponds to the power of the collected electrical carriers. Of course, at every given wavelength, each involved phenomenon is connected to the others. For example, reducing the flux leaving the cell to lower the optical losses may increase the collected power, but may increase the heat source as well. This question needs to be raised at the level of the full solar spectrum. The Sankey diagram of the transformation of the total incident solar power (Figure 8) shows that there is no guarantee -a priori- that minimizing the optical and electrical losses will always maximize the power converted into electricity: thermal losses may also increase.

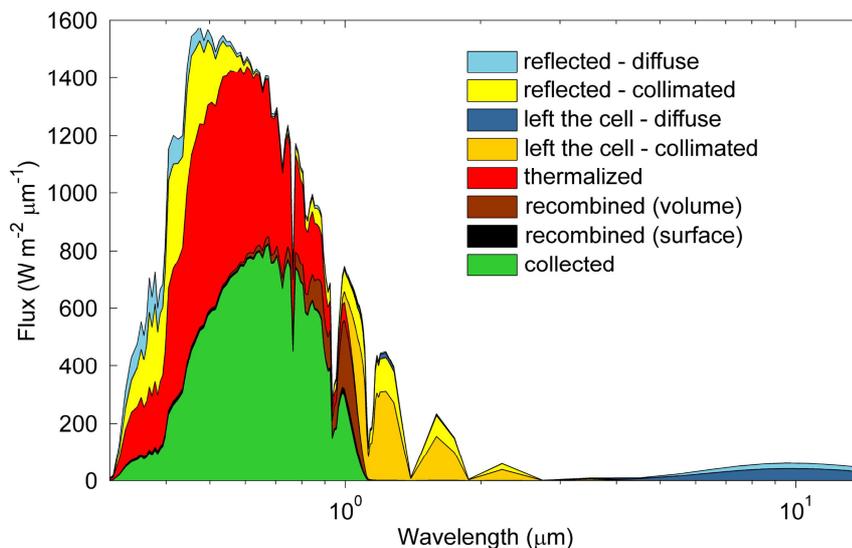

Fig. 7. Spectral distribution of the incident irradiation (AM1.5G solar spectrum and surroundings).

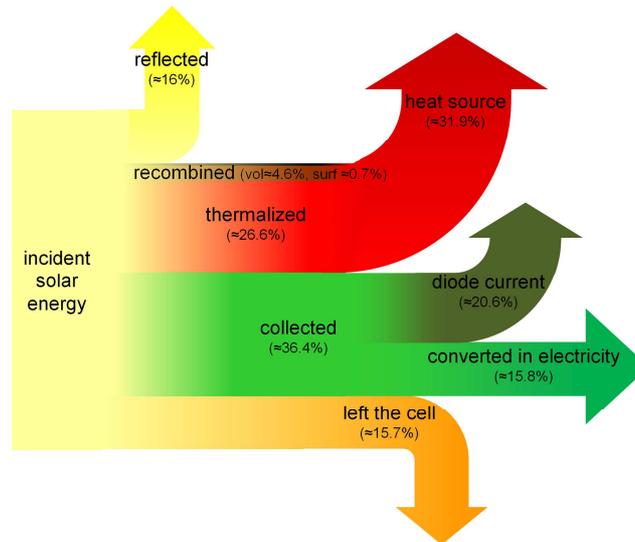

Fig. 8. Sankey diagram of the conversion of the incident power.

### 3) Handling of diffuse irradiations

The standard solar spectrum normally used in PV simulations is called AM1.5G where G stands for global and which includes both the direct and diffuse components of the solar irradiation. However, the multi-directional nature of the diffuse part is often ignored in the modeling of radiative transfer since it makes it somehow complicated. TASC-1D can handle direct and diffuse solar irradiations as well as any kind of additional diffuse external irradiation.

As a result, not only the direct and diffuse components are described correctly in terms of radiative transfer within the cell and at the interfaces, but one can track the transformations of these fluxes. Figure 7 highlights that if the antireflection coating is optimized for a normal direct illumination at 0.65 µm, it reflects a fraction of the diffuse component at all wavelengths almost identically. As for the infrared region, it should be noticed that the diffuse flux leaving the cell includes the parts transmitted and emitted by the cell. The difference between the flux absorbed by the cell and the one thermally emitted gives rise to a radiative heat source. Depending on the presence of an external irradiation, on the outcome of a difference between the cell and the surrounding medium temperatures when the heat transfer problem is solved, this part of the electromagnetic spectrum, usually neglected, is likely to play a role in the cooling or heating of the cell. However, the resulting thermal behavior should be taken with caution since no encapsulation is considered yet. In order to understand better how TASC-1D can help in scrutinizing the thermal impacts on PV cell performances, the next section is dedicated to the analysis of these impacts when the temperature of the cell is not prescribed, but driven by the thermal and radiative BCs and the cell characteristics.

### C. Analysis of the thermal impacts when the heat transfer problem is solved

Analyses were made so far in prescribed thermal states: the temperature was fixed. While they exhibited the spectral variations of some of the heat sources generated within the cell, they did not give any information on how the resulting thermal behavior would impact the PV cell performances.

A selection of scenarios, summarized in Table 2, was chosen to demonstrate the ability of TASC-1D to determine the variations of the cell operating temperature and performance as a function of the thermal boundary conditions. Cases 1, 2 and 3 consider several outdoor temperatures to simulate different locations or seasonal variations at the same location. Those cases are splitted into three sub-cases to evaluate the effect of the heat transfer coefficients. The indices a, b and c are respectively cases with no wind, a light breeze (about 8 km.h$^{-1}$) and some wind (about 20 km.h$^{-1}$) [8]. Case 4 models a device that cannot evacuate heat by its back surface (such as some building integrated PV panels for instance). Case 5 represents a PV device cooled on its rear side (such as a hybrid solar panel). All simulations were made using the AM1.5G spectrum with its diffuse (100.4 W.m$^{-2}$) and direct (901.2 W.m$^{-2}$) components.

Table 2
SIMULATION CASES

| Case | External temperatures (°C) front/back | Heat transfer coefficients (W.m-2.K-1) front/back |
|------|----------------------------------------|----------------------------------------------------|
| 1.a | 10/10 | 5/5 |
| 1.b | 10/10 | 14/5 |
| 1.c | 10/10 | 23/5 |
| 2.a | 25/25 | 5/5 |
| 2.b | 25/25 | 14/5 |
| 2.c | 25/25 | 23/5 |
| 3.a | 35/35 | 5/5 |
| 3.b | 35/35 | 14/5 |
| 3.c | 35/35 | 23/5 |
| 4 | 25/25 | 5/0 |
| 5 | 25/25 | 5/100 |

The resulting efficiencies and cell equilibrium temperatures are given in Figure 9. The noticeable trends are fairly straightforward. Efficiency varies contrarily to the cell equilibrium temperature. Since the incoming power is prescribed, this temperature is determined by the heat transfer coefficient between the cell and its surroundings and by the surrounding temperatures. The important impact on efficiency of the wind condition, well known by the PV community and modeled here by the heat transfer coefficients, appears clearly. The order of magnitude of the calculated cell temperatures is consistent with the values found in the literature. However fully meaningful comparisons will only be possible when the encapsulation of the cell is added to the code.

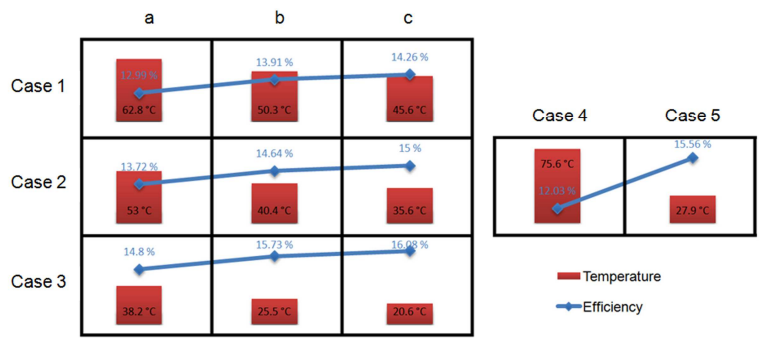

Fig. 9. Efficiency and equilibrium temperature of the different scenarios.

Results for cases 2.a, 4 and 5 indicate how important the panel configuration is. The comparison of the efficiency in case 4 with the one of a classic "open rack" PV installation (case 2.a) reminds one of the main challenges for integrated photovoltaics. Also, one can remember Figure 6 and have a look at the results in case 5 to acknowledge the potentiality of generating heat together with electricity. However, it is not so simple since some power and equipments will be required in order to properly harvest this thermal source (electrical power and pumps to circulate a cooling fluid for example).

This analysis shows that TASC-1D-cSi can predict the cell temperature resulting from the balance between the total heat source generated within the cell and its evacuation towards the environment as a function of the thermal BCs. The heat sources do not simply derive from the incoming solar flux but also are function of the cell temperature: there is an intricate connection among optical-radiative, electrical and thermal responses of the system. Beyond searching for a smart cooling system that would reduce the temperature of the PV cell and beyond reducing the optical and electrical losses by adjusting the cell characteristics, this simulation tool allows exploring new possibilities for optimizing PV cells. For example, we could introduce a thermal criterion such as minimizing the total heat source generated within the cell.

## IV. CONCLUSION

In this study, we have presented TASC-1D-cSi, an original simulation tool allowing in-depth analyses of the thermal losses in crystalline silicon solar cells. The 1D modeling includes the combined mechanisms of radiative transfer in the visible and partly in the infrared, of electrical carrier transport, and of heat transfer. In the current version of the code, a PN homo-junction without encapsulation but possibly with an antireflection coating is subjected to direct and diffuse solar irradiations and to an infrared diffuse isotropic radiation source from the surrounding. The cell thermal equilibrium is driven by the heat sources generated within it and by exchanges with the surrounding through various kinds of boundary conditions. We have summarized the main elements of the modeling and the solution methods employed. With this tool, the cell characteristics, the radiative and thermal boundary conditions can be varied and as a results, a large set of outputs tell about the thermal losses that reduce the cell performances, in addition to and in connection with the optical and electrical losses. After validations of some of the outputs against results from PC1D and existing data in the literature, we have highlighted the capabilities of the code. It allows thorough analyses, for example of the quantum efficiencies, of the optical – electrical – thermal power loss mechanisms, and considers properly the direct and diffuse components of irradiations. More importantly, we have shown how the code, by solving the heat transfer problem, provides new detailed information about the performances of cSi PV cells as a function of the thermal conditions, but also of the characteristics of the cell. It has been concluded that taking into account the thermal losses when designing PV cells would be likely to provide conclusions different from those obtained when only the optical and electrical losses are taken into consideration. Including a thermal criterion in the optimal design of PV cells may give rise to new, possibly unexpected, efficient engineering solutions. Before doing so, the detailed mechanisms that rule the cell conversion efficiency as a function of temperature, and at the same time, that also govern the heat sources, must be understood better. TASC-1D-cSi can help for all these purposes. Lastly, it should be mentioned that the code was initially developed for the case of standard crystalline silicon PN junctions because there is a lot of data available about the electrical, optical, and thermal properties of the materials. In the future, the simulation tool will be extended by considering the encapsulation layers and the back contacts and by making the modifications required when the cell is subjected to concentrated solar irradiations. Other kinds of PV cells could be examined in the same manner so as to understand the thermal impacts on their performances, as it was already done in the case of nano-thermophotovoltaic cells [9].

APPENDIX

Anti-reflection coatings (ARC) consist of one or several thin layers of dielectric materials having appropriate optical properties. These layers are deposited or grown onto the cell to lower the reflection of light in the spectral range where the photo-conversion takes place. In the version of the code presented in [7], the calculated photoelectric efficiencies were small in comparison with the usual order of magnitude of commercial silicon PV cells because no ARC was considered. The code can now handle an ARC of several layers of the following materials: SiN, $SiO_2$, $TiO_x$, ZnS. Calculation of the radiative properties (reflectance, transmittance) of the whole coating is made as a function of frequency, angle of incidence, layer thickness and complex refractive index of the materials. The latter are taken from the work of Nagel et al. [10]. The theory for reflectance calculation of multilayer thin films is available in the book of Born & Wolf [11]. Our calculations results were validated against data reckoned with the IMD software developed by Windt [12].

The intrinsic carrier concentration plays a central role in the calculation of electrical properties. It was initially chosen to use the theoretical expression which involves the carrier effective densities of states in the conduction and valence bands, as well as the fundamental indirect band gap of silicon and its variation with temperature (*T*). But it was observed that for silicon the expression from Sproul & Green [13] is in better agreement with experimental data in the 275-375 K temperature range. By comparison of results from this formula and the one we employed previously, we have observed that despite the temperature exponent difference, the variations with temperature are very similar over the range of interest. As the intrinsic carrier concentration fits well with other experimental data around room temperature [14] we have decided to adopt this expression.

The mobility of the carriers depends on the dopant concentrations and on the temperature and also on whether the carriers are minority or majority ones. In this frame, we have rethought the choice of the expressions for the minority carrier mobilities. We have selected the Klaassen unified mobility model [15] that is widely used in the PV community and distinguishes between majority and minority carriers. A review from Altermatt [6] compares Klaassen's model to the different existing experimental values for holes and electrons mobilities and concludes that the match is fairly good.

The expression used to calculate the hole lifetime in N-doped region was taken from Cuevas et al. [16] and gives 0.0416 µs at the selected dopant concentration. Implementing an expression that distinguishes the Shockley-Read-Hall, Auger and radiative processes and includes their temperature dependences is currently being considered.

Eventually, we provide some errata to the previous paper introducing the simulation tool TASC-1D [7]: the parameter $A_d$ in the expression of the interband absorption coefficient was incorrectly copied (in Table 1) from [17]; the correct value is 1.052 $10^6$ instead of 1.052 $10^{-6}$. The surface recombination term was added while it ought to be subtracted in the thermal boundary condition and the thermal balance. In the expression of the free carrier absorption coefficient ($\kappa_{fc}$) from Clugston & Basore [18] (equation 29) $10^4$ should be replaced by $10^{-4}$.